\newif \ifdraft \drafttrue

\documentclass[10pt,preprint,nocopyrightspace]{sig-alternate-10pt} 

\usepackage{xcolor}
\usepackage[lowtilde]{url}
\usepackage{alltt}
\usepackage{booktabs}
\usepackage{graphicx}
\usepackage{tikz}
\usepackage{multirow}
\usepackage{xspace}
\usepackage{caption}
\DeclareCaptionType{copyrightbox}
\usepackage{subcaption}
\usepackage{balance}

\ifdraft
\newcommand{\todo}[1]{\textcolor{red}{TODO: #1}}
\newcommand{\ks}[1]{\textcolor{orange}{ks: #1}}
\newcommand{\mwh}[1]{\textcolor{blue}{mwh: #1}}
\newcommand{\lv}[1]{\textcolor{red}{lv: #1}}
\newcommand{\jnf}[1]{\textcolor{purple}{jnf: #1}}
\else
\newcommand{\todo}[1]{}
\newcommand{\mwh}[1]{}
\newcommand{\ks}[1]{}
\newcommand{\lv}[1]{}
\newcommand{\jnf}[1]{}
\fi

\newcommand{\boom}[1]{\textsc{#1}}
\newcommand{\fwzero}{\boom{firewall}$\leftharpoondown$\xspace}
\newcommand{\fwone}{\boom{firewall}$\rightleftharpoons$\xspace}
\newcommand{\fwtwo}{\boom{firewall}$\rightleftharpoons\!\!\odot$\xspace}

\begin{document}

\newcommand{\name}{Morpheus\xspace}
\newcommand{\Name}{MORPHEUS\xspace}

\title{\name: Safe and Flexible Dynamic Updates for SDNs}

\numberofauthors{3} 
\author{
\alignauthor Karla Saur\\
        \affaddr{University of Maryland}\\
        \email{ksaur@cs.umd.edu}
\alignauthor Joseph Collard\\
        \affaddr{UMass Amherst}\\
        \email{jcollard@cs.umass.edu}
\alignauthor Nate Foster\\
        \affaddr{Cornell University}\\
        \email{jnfoster@cs.cornell.edu}
\and
\alignauthor Arjun Guha\\
        \affaddr{UMass Amherst}\\
        \email{arjun@cs.umass.edu}
\alignauthor Laurent Vanbever\\
        \affaddr{ETH Zurich}\\
        \email{lvanbever@ethz.ch}
\alignauthor Michael Hicks\\
        \affaddr{University of Maryland}\\
        \email{mwh@cs.umd.edu}
}
\maketitle

\begin{abstract}
SDN controllers must be periodically modified to add features, improve
performance, and fix bugs, but current techniques for implementing
dynamic updates are inadequate. Simply halting old controllers and
bringing up new ones can cause state to be lost, which often leads to
incorrect behavior---e.g., if the state represents hosts blacklisted
by a firewall, then traffic that should be blocked may be allowed to
pass through. Techniques based on record and replay can reconstruct
state automatically, but they are expensive to deploy and can lead to
incorrect behavior. Problematic scenarios are especially likely to
arise in distributed controllers and with semantics-altering updates.

This paper presents a new approach to implementing dynamic controller
updates based on explicit state transfer. Instead of attempting to
infer state changes automatically---an approach that is expensive and
fundamentally incomplete---our framework gives programmers effective
tools for implementing correct updates that avoid major disruptions.
We develop primitives that enable programmers to directly (and easily,
in most cases) initialize the new controller's state as a function of
old state and we design protocols that ensure consistent behavior
during the transition. We also present a prototype implementation
called \name, and evaluate its effectiveness on representative case
studies.
\end{abstract}



\section{Introduction}

Software-defined networking (SDN) controllers are complex software
systems that must simultaneously implement a range of interacting
services such as topology discovery, routing, traffic monitoring, load
balancing, authentication, access control, and many others. Like any
non-trivial system, SDN controllers must be periodically updated to
add features, improve performance, and fix bugs. However, in many
networks, downtime is unacceptable, so updates must be
deployed \emph{dynamically}, while the network is in operation and in
such a way as to minimize disruptions.

%
%
In general, dynamic updates differ from static ones in that while
modifying the \emph{program code} they must also be concerned with the
current \emph{execution state}. In SDN, this state can be divided into
the \emph{internal state} stored on controllers (e.g., in memory, file
systems, or databases), and the \emph{external state} stored on
switches (e.g., in forwarding rules). A key challenge is that updated
code may make different assumptions about state---e.g., using
different formats to represent internal data structures or installing
different  rules on switches. This challenge is exacerbated
in SDN, where state does not reside at a single location but is
instead distributed across multiple controllers and switches.

%
%
\paragraph*{Existing approaches}
Most SDN controllers today employ one of two strategies for performing
dynamic updates, distinguished by how they attempt to ensure correct,
post-update execution state.

\noindent
$\bullet$ In \emph{simple restart}, the system halts the old
controller and begins executing a fresh copy of the new controller. In
doing so, the internal state of the old controller is discarded
(except for persistent state---e.g., stored in a database), under the
assumption that any necessary state can be reconstructed by the new
controller after the restart. One simple strategy for recovering
internal state is to delete the forwarding rules installed on switches
so future network events are sent to the controller, which can use
them to populate its state. This behavior is available by default in
open-source SDN platforms such as POX~\cite{pox} and
Floodlight~\cite{floodlight}.


\noindent $\bullet$ In \emph{record and replay}, the system 
maintains a trace of network events received by the old
controller. During an update, the system first replays the logged
events to the new controller to ``warm up'' its internal state and
then swaps in the new controller for the old one. By giving the new
controller access to the events that were used to generate the
internal and external state for the old controller, it is possible to
avoid the issues that arise with less direct mechanisms for
reconstructing state. Record and replay has been used effectively in
several previous systems including HotSwap~\cite{hotswap},
OpenNF~\cite{opennf}, and a recent system for managing
middleboxes~\cite{rrmbox}. A related approach is to attempt
to reconstruct the state from the existing forwarding rules on the
switches, rather than from a log. According to private discussions
with SDN operators, this approach is often adapted by proactive
controllers that do not make frequent changes to network state (e.g.,
destination-based forwarding along shortest paths).

%

Unfortunately, neither approach constitutes a general-purpose solution
to the dynamic update problem: they offer little control over how
state is reconstructed and can impose excessive performance
penalties. Simple restarts discard internal state, which can be
expensive or impossible to reproduce. In addition, there is no
guarantee that the reconstructed state will be harmonious with the
assumptions being made by end hosts---i.e, existing flows may be
routed along different paths or even to different destinations,
breaking socket connections in applications. Record and replay can
reproduce a harmonious state, but requires a complex logging system
that can be expensive to run, and still provides no guarantees about
correctness---e.g., in cases where the new controllers would have
generated a different set of events than the old controller
did. Reconstructing controller state from existing forwarding rules
can be laborious and error prone, and is risky in the face of
inevitable switch failures. We illustrate these issues using examples
in Section~\ref{sec:problem}.

\paragraph*{Our approach: Update by state transfer}
The techniques just discussed \emph{indirectly} update network state
after an update. This paper proposes a more general and flexible
alternative: to properly support dynamic updates, operators should be
able to \emph{directly} update the internal state of the controller,
as a function of its current state. We call this idea \emph{dynamic
update by state transfer}.

To support this dynamic update technique, controllers must offer three
features: (1) they need a way of making their internal state
available; (2) they need a way of initializing a new controller's
state, starting from (a possibly transformed version of) the old
controller's state; and (3) they need a way to coordinate behavior
across components when updates happen, to make sure that the update
process yields a consistent result. This basic approach has been
advocated in prior work on \emph{dynamic software updates}, which has
shown that these requirements are relatively easy to
meet~\cite{kitsune,rubah}.

Update by state transfer directly addresses the performance and
correctness problems of prior approaches. There is no need to log
events, and there is no need to process many events at the controller,
whether by replaying old events or by inducing the delivery of new
ones by wiping rules. Moreover, the operator has complete control over
the post-update network state, ensuring that it is harmonious with the
network---e.g., preserving existing flows and policies. The main costs
are that the network programmer must write a function (we call it
$\mu$) to initialize the new controller state, given the old
controller state, and the controller platform must provide protocols
for coordinating across distributed nodes.

Fortunately, our experience (and that of dynamic software updates
generally~\cite{kitsune,rubah}) is that for real-world updates this
function is not difficult to write and can often be partially
automated. The changes to the controller platform needed to support
state initialization and coordination between nodes adds complexity,
but they are one-time changes and are not difficult to
implement. Moreover, the cost of coordinating controller nodes is
likely to be reasonable, since the distributed nodes that make up the
controller are likely to be relatively small and either co-located on
the same machine or connected through a fast, mostly reliable network.

\paragraph*{\name: A controller with update by state transfer}
We have implemented our approach in \name, a new distributed
controller platform based on
Frenetic~\cite{frenetic,netkat,consistentupdate}, described in
Section~\ref{sec:boombox}. Our design is based on a distributed
architecture similar to the one used in industrial controllers such as
Onix~\cite{koponen2010onix} and ONOS~\cite{onos}. We considered
modifying a simpler open-source controller such as POX or
Floodlight~\cite{pox,floodlight}, but decided to build a new
distributed controller to provide evidence that update by state
transfer will work in industrial settings.

\name employs a NIB, basic controller replicas, and standard 
applications for computing and installing forwarding paths, each
running as separate applications. Persistent internal state is stored
in the NIB, which can be accessed by any application. When an
application starts (or restarts, e.g., after a crash) it connects to
the NIB to access the state it needs, and publishes updates to the
state while it runs. Applications coordinate rule deployments to
switches via controller replicas, which use NetKAT~\cite{netkat} to
combine policies into a unified policy, and can use consistent
updates~\cite{consistentupdate} to push rules to switches.

Supporting update by state transfer requires only a few additions to
\name's basic design, described in Section~\ref{sec:updates}. The
relevant state is already available for modification in the NIB, so we
just need a means of modifying that state to work with the new
versions. 

We also need to coordinate the update across the affected
applications. To see why this is important, consider a situation in
which we have several routing application replicas, each responsible
for a subset of the overall collection of switches.
Now suppose we wish to deploy a dynamic update that changes which
paths forward traffic through the network. It is clear we must update
all of the replicas in a coordinated manner, or else some of the
replicas could implement old paths and others implement new paths,
leading to anomalies including loops, black holes, etc. \name's simple
coordination protocol operates in three steps:
(1) \emph{quiescence}---the affected applications are signaled and
paused before the update begins; (2) \emph{installation}---the $\mu$
function is registered with the NIB for the purposes of transforming
the state; and (3) \emph{restart}---the updated applications are
restarted, using $\mu$ to update the NIB state (in a coordinated
way). After the state is updated, they send updated policies to the
controller replicas which compose them and generate rules to install
on switches.

Using \name we have written several applications, and several versions
of each, including a stateful firewall, topology discovery, routing,
and load balancing. Through a series of experiments, described in
Section~\ref{sec:exps}, we demonstrate the advantages of update by
state transfer, compared to simple restarts and record-and-replay. In
essence, there is far less disruption, and no incorrect behavior. We
also find that the $\mu$ functions are relatively simple, and an
investigation of changes to open-source controllers suggests that
$\mu$ functions for realistic application evolutions would be simple
as well.

\paragraph*{Summary}
This paper's contributions are as follows:
\begin{itemize}
\setlength{\itemsep}{0pt}
\item We study the problem of performing dynamic updates to SDN controllers 
and identify fundamental limitations of current approaches.
\item We propose a new, general-purpose solution to dynamic update
  problem for SDNs---\emph{dynamic update by state
    transfer}. With this solution, the programmer explicitly
  transforms old state to be used with the new controller, and an
  accompanying protocol coordinates the update across distributed nodes.
\item We describe a prototype implementation of these ideas in the \name system.
\item We present several applications as case studies as well as experiments 
showing that \name implements updates correctly and with far less
disruption than current approaches.
\end{itemize}
Next, we present the design and implementation of \name
(\S\ref{sec:problem}-\ref{sec:updates}), our experimental evaluation
(\S\ref{sec:exps}), and a discussion of related work and conclusion
(\S\ref{sec:related_work}-\ref{sec:conclusion}).
 
\section{Overview}
\label{sec:problem}


This section explains why existing approaches for handling dynamic
updates to SDN controllers are inadequate in general, and provides
detailed motivation for our approach based on \emph{state transfer}.

\begin{figure}[t]
\center\tikz\node{\pgfimage[width=.7\columnwidth]{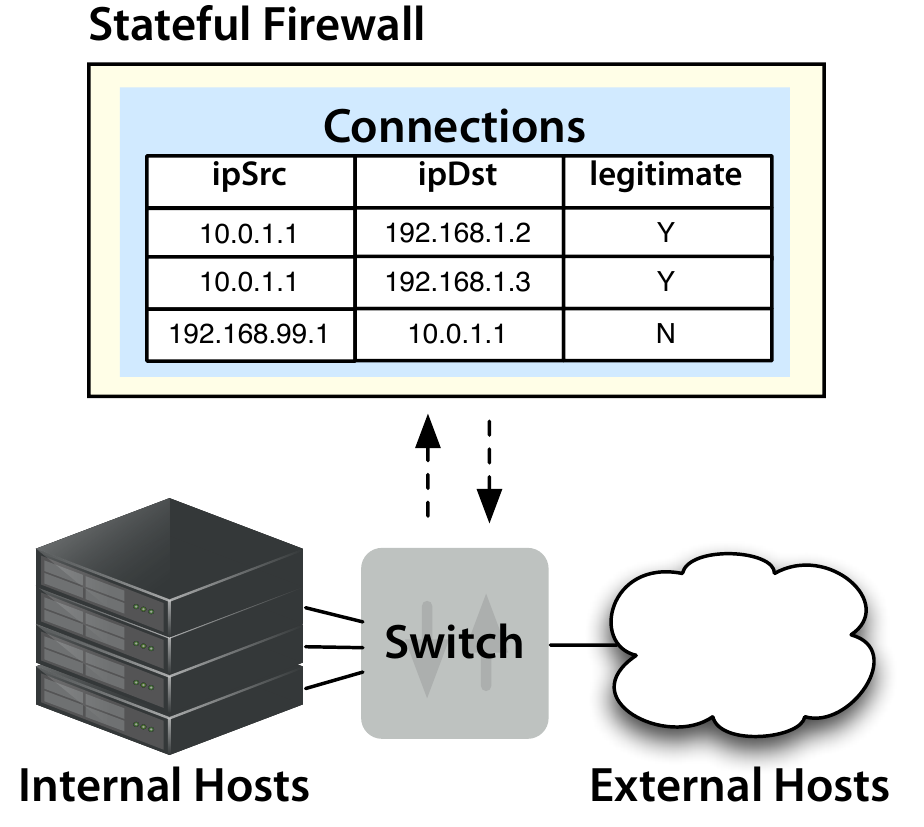}};
\caption{Example application: stateful firewall.}
\label{fig:firewall}
\end{figure}

\subsection{Simple Restart}

As an example, suppose the SDN controller implements a stateful
firewall, as depicted in Figure~\ref{fig:firewall}. The topology
consists of a single switch connected to trusted internal hosts and
untrusted external hosts. Initially the switch has no forwarding
rules, so it diverts all packets to the controller. When the
controller receives a packet from a trusted internal host, it records
the internal-external host pair in its (internal) state and punches a
hole in the firewall so the hosts can communicate. Conversely, if the
controller receives a packet from an external host first, it logs the
connection attempt and drops the packet.

Now suppose the programmer wishes to update the firewall so that if an
external host tries to initiate more than $n$ connections, then it is
blacklisted from all future communications.  With simple restart, the
old controller would be swapped out for a new controller that contains
no record of connections initiated by internal and external hosts. In
the absence of any other information about the state prior to the
update, the controller would delete the rules installed on the switch
to match its own internal state, which is empty. This leads to a
correctness problem:\footnote{It may also be disruptive: if unmatched
traffic is sent to the controller, then the new controller will
essentially induce a DDoS attack against itself as a flood of packets
stream in.}  If the external host of an active connection sends the
first few packets after the rules are wiped, then those packets will
be interpreted as unsolicited connection attempts. The host could
easily be blacklisted even though it is merely participating in a
connection initiated previously by an internal host.

This problem stems from the fact that the old controller's state is
discarded by the simple restart. In this example, it could be avoided
by storing key internal state outside of the controller process's
memory---e.g., in a separate \emph{network information base} (NIB), as
is done in controllers such as Onix~\cite{koponen2010onix}---and
indeed, we do exactly this in our \name controller. However, in
general, safe dynamic updates require more than externalized state, as
we discuss in Section~\ref{sec:dsu}---e.g., in the case that the new
version expects the state in a new format and multiple controllers
share this state.



\begin{figure}[t]
\center\tikz\node{\pgfimage[width=.7\columnwidth]{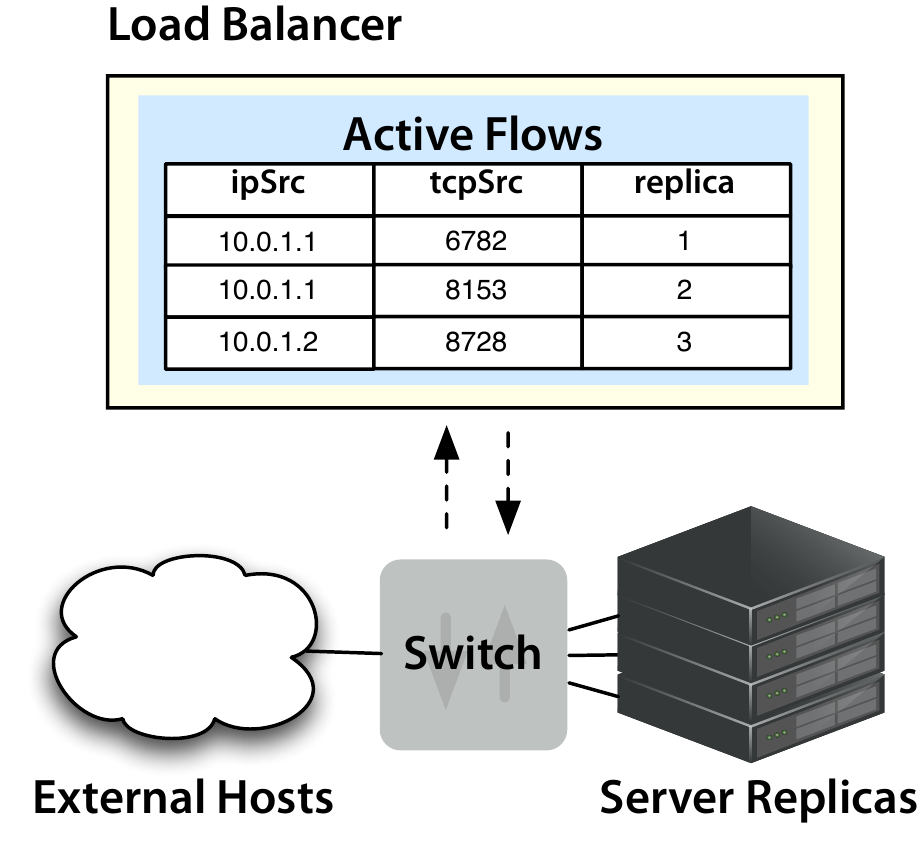}};
\caption{Example application: load balancer.}
\label{fig:loadbalance}
\end{figure}

\subsection{Record and Replay} \label{sec:rr}

At first glance, record and replay seems like it might offer a fully
automatic solution to dynamic controller updates. The HotSwap (HS)
system~\cite{hotswap}, a noteworthy example of this approach, records
a trace of the events received by the old controller and replays them
to the new controller to ``warm up'' its internal state before
swapping it in. For the stateful firewall, HS would replay the network
events for each connection initiated by an internal host and so would
easily reconstruct the set of existing connections, avoiding the
problems with the simple restart approach.  Moreover, because record
and replay works with events drawn from a standard API like OpenFlow,
it is fully ``black box''---the implementation details of the old and
new controllers are immaterial.

Unfortunately, record and replay has several limitations that prevent
it from being a full solution to the dynamic update problem. One
obvious issue is overhead: in general, unless the system has prior
knowledge of the new controller's functionality (which it will not, in
general), the system will have to record (and replay) all relevant
events that contributed to the network's current state. Doing this can
be prohibitively expensive in a large, long-running network.

Another issue is that the recorded trace may not make sense for the
new controller, and therefore replaying it may result in an incorrect
state, for the following reasons. The new controller may, in general,
behave differently than the old one---e.g., it may install different
forwarding rules on switches. As such, if the new controller had been
used from the start, these rules might have caused a different set of
network events to be generated than those that were actually
recorded. Such events could have been induced \emph{directly} due to
different rules (e.g., because they handle fewer or more packets
compared to the old rules) or they might have been
induced \emph{indirectly} (e.g., because the new rules elicit
different responses from the hosts that are communicating via the
network). Establishing that an update is correct under these
circumstances is extremely difficult in general.

To illustrate, consider the example of a server load balancer, as
depicted in Figure~\ref{fig:loadbalance}. The topology consists of a
single switch with one port connected to a network of external hosts
and another $n$ ports connected to back-end server
replicas. Initially, the switch has no rules, so all packets are
diverted to the controller. Upon receiving a new connection from an
external host, the controller picks a server replica (e.g., uniformly
at random) and installs rules that forward traffic in both directions
between the host and the selected server. The controller also records
the selected server in its internal state (e.g., so it can correctly
repopulate the forwarding rules if the switch drops out and
later reconnects).

Now suppose the programmer wishes to dynamically deploy a new version
of the controller where the selection function selects the least
loaded server and also puts a cap $c$ on the number of open
connections to any given server, and refuses connections that would
cause a servers to exceed that cap. During replay, the new controller
would receive a network event for each existing connection
request. However, it would remap those connections to the least loaded
server instead of the server previously selected by the old
controller. In general, the discrepancy between these two load
balancing strategies will break connection affinity---a different
server replica may receive the $i$th packet in a flow and reset the
connection.


Attempting to reconstruct the controller state from querying the
switch state could also be problematic. Although the new controller
would have the information needed to generate forwarding rules that
preserve connection affinity, writing the controller to retrieve this
information is potentially laborious, error-prone work for the
programmer. And it may require modifications to the code; e.g., if the
new controller uses statically allocated data structures to keep track
of active flows (something that is possible due to the cap $c$), it
may be incorrect to exceed the cap. We would prefer a solution that is
simpler and more systematic.

\subsection{Solution: Update by state transfer}
\label{sec:dsu}

This paper proposes a different approach to the SDN dynamic update
problem. Rather than attempting to develop fully automated solutions
that handle certain simple cases but are more awkward or impossible in
others, we propose a general-purpose solution that attacks the
fundamental issue: \emph{dynamically updating the state.}  The above
approaches attempt to \emph{indirectly} reconstruct a reasonable
state, but they lack sufficient precision and performance to fully
solve the problem.

Our approach, which we call \emph{update by state transfer}, solves
the dynamic update problem by giving the programmer \emph{direct}
access to the running controller's state, call it $\sigma$, along with
a way enabling the new controller with an existing state, call it
$\sigma'$, such that the new state can be constructed as a function,
call it $\mu$, of the old state so that $\sigma' = \mu(\sigma)$.  In
addition, our approach requires a means to signal the controller that
an update is available so that it can \emph{quiesce} prior to
performing the update. This mechanism ensures that $\sigma$ is
consistent (e.g., is not in the middle of being changed) before using
$\mu$ to compute $\sigma'$.

Consider the problematic examples presented thus far. For both the
firewall update and the load balancing update, the state transfer
approach is trivial and effective: setting the $\mu$ function to a
no-op (i.e., identity function) grandfathers in existing connections
and the new semantics is applied to new connections. Pleasantly, for
the load-balancing update, any newly added replicas will receive all
new connection requests until the load balances out.

Another feature of update by state transfer is that it permits the
developer to more easily address updates that are
backward-incompatible, such as the load balancer with a cap $c$
discussed above. In these situations, the current network conditions
may not constitute ones that could ever be reached had the new
controller been started from scratch. With state transfer, the
operator can either allow this situation temporarily by preserving the
existing state, with the new policy effectively enforced once the
number goes below the cap. Or she can choose to kill some connections,
to immediately respect the cap. The choice is hers. By contrast, prior
approaches will have unpredictable effects: some connections may be
reset while others may be inadvertently grandfathered in, unbeknownst
to the controller.

In addition to its expressiveness benefits, update by state transfer
has benefits to performance: it adds no overhead to normal operation
(no logging), and far less disruption at update-time (only the time to
quiesce the controller and update the state). The main cost is that
the network service developer needs to write $\mu$, which will not
always be a no-op. For example, if we updated a routing algorithm
from using link counts to using current bandwidth measurements, the
controller state would have to change to include this additional
state. Fortunately, according to our experience (and that of a
substantial body of work in the related area of \emph{dynamic software
updating}), $\mu$ tends to be relatively simple, and its construction
can be at least partially automated.

\section{\Name Controller}
\label{sec:boombox}


\begin{figure}
\centering
  \includegraphics[scale=0.65]{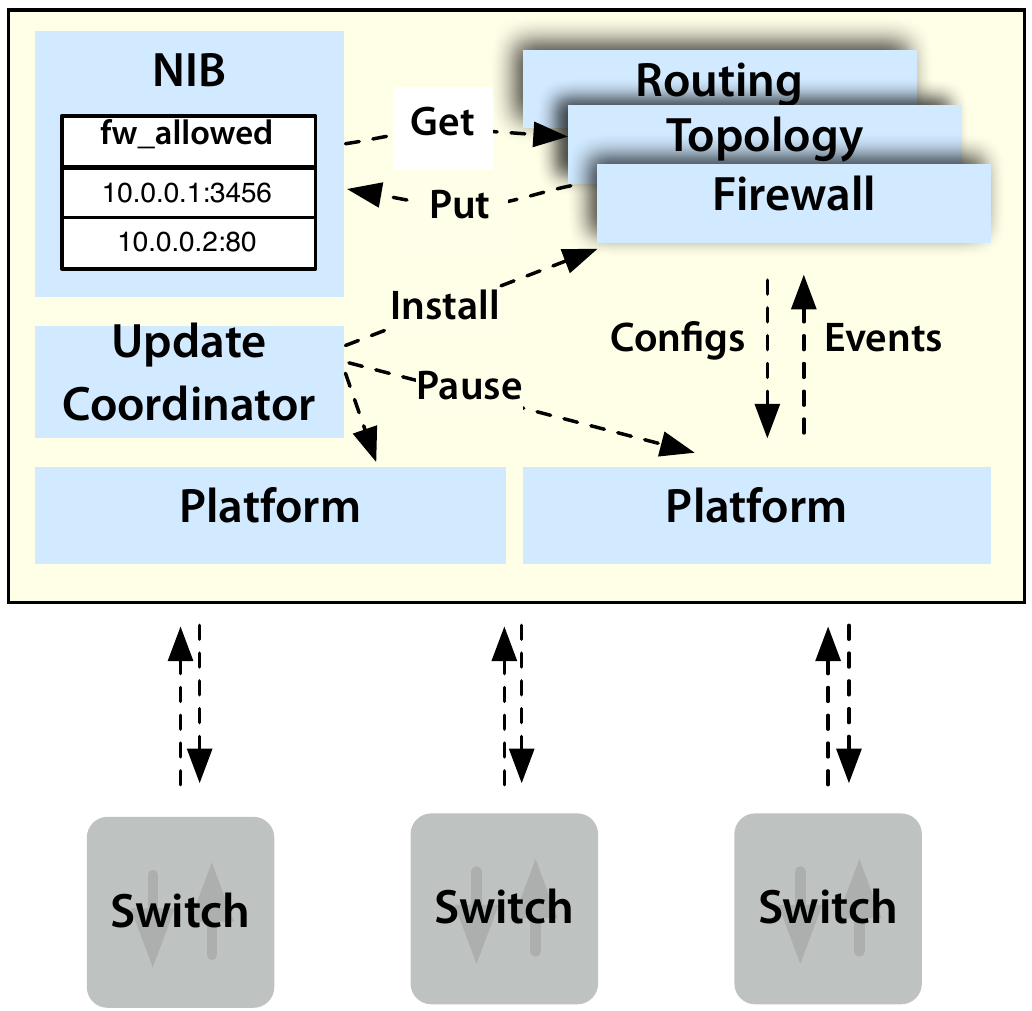}
\caption{\name architecture.}
\label{fig:boombox}
\end{figure}

To provide a concrete setting for experimenting with dynamic SDN
updates, we have implemented a new distributed controller
called \name, implemented in Python and based on the Frenetic
libraries~\cite{frenetic,netkat,consistentupdate}. Our design follows
the basic structure used in a number of industrial controllers
including Onix~\cite{koponen2010onix} and ONOS~\cite{onos}, but adds a
number of features designed to facilitate staging and deployment of
live updates. This means that it should be easy to adapt update
techniques developed in the context of \name for use in other
controllers as well. We should note that our aim is to support updates
to the applications running on the controller, but not necessarily the
controller itself. In the future, we plan to investigate extensions
that will also support updates to the controller
infrastructure---e.g., migrating to a new protocol for communicating
with SDN switches.

\subsection{Architecture}
\name's architecture is shown in Figure~\ref{fig:boombox}. 
The controller is structured as a distributed system in which
nodes communicate via well-defined message-passing interfaces.
\name provides four types of nodes:
\begin{itemize}
\item{} platform nodes (\boom{platform}), which are responsible for 
managing low-level interactions with SDN switches and interfacing with
applications,
\item{} a network information base (\boom{nib}), which provides 
persistent storage for application state,
\item{} an update coordinator (\boom{updc}), which implements distributed 
protocols for staging and deploying updates, and
\item{} application nodes (\boom{topology}, \boom{routing}, etc.), which 
implement specific kinds of functionality, such as discovering
topology or computing shortest paths through the topology.
\end{itemize}
Each node executes as a separate OS-level process, supporting
concurrent execution and isolation.
Processes also make it easy to use existing OS tools
to safely spawn, execute, replicate, kill, and restart nodes.

\subsection{Components}

We now describe \name's components in detail.

\paragraph*{Platform}
The most basic components are \boom{platform} nodes, which implement
basic controller functionality: accepting connections from switches,
negotiating features, responding to keep-alive messages, installing
forwarding rules, etc. The \boom{platform} nodes implement a simple
interface that provides commands for interacting with switches:
\begin{itemize}
\item \texttt{event()} returns the next network event,
\item \texttt{update(pol)} sets the network configuration to \texttt{pol}, specified using NetKAT~\cite{netkat},
\item \texttt{pkt\_out(sw,pkt,pt)} injects packet \texttt{pkt} into the network at \texttt{sw} and \texttt{pt}, 
\end{itemize}
as well as commands for synchronizing with the \boom{updc} during dynamic updates:
\begin{itemize}
\item \texttt{pause()} temporarily stops propagating configurations to the network, and
\item \texttt{resume()} resumes propagating configurations.
\end{itemize}
When multiple \name applications are operating, the \boom{platform}
nodes make every network event available to each application by default.
If needed, filtering can be applied to prevent some applications from seeing
some network events. Likewise, the policies provided by each application
are combined into a single network-wide policy using NetKAT's modular
composition operators~\cite{netkat}. For scalability and fault tolerance,
\name would typically use several \boom{platform} nodes that each
manage a subset of the switches. These nodes would communicate with each
other to merge their separate event streams into a single stream, and
similarly for NetKAT policies. For simplicity, our current implementation
uses a single \boom{platform} node to manage all of the switches
in the network.

\paragraph*{Network Information Base}
\name applications store persistent state in the \boom{nib}. 
The information in the \boom{nib} is guaranteed to be preserved across
application executions, thereby avoiding disruption if individual
applications stop and restart. The
\boom{nib} provides a simple interface to a NoSQL key-value store,
which can be used to store persistent state.\footnote{Obviously,
  applications may also maintain their own in-memory state for
  efficiency reasons, but this state is lost on restart.}
\name's store is currently based on
Redis~\cite{redis} and although it currently uses a single node, Redis
supports clustering for better scalability.

Data stored in the NIB is divided among conceptual \emph{namespaces},
organized according to the applications that use it. For
example, a firewall application might store information in the
\boom{nib} in the \texttt{fw\_allowed} namespace 
about which hosts are currently allowed. An application may access data in multiple
namespaces, where it might be the conceptual data owner for one, but a consumer
of another. For example, our \boom{topology} application discovers the
structure of the network by interacting with the \boom{platform}
nodes, and stores the topology persistently in
the \texttt{topology} namespace. 

Redis does not 
support namespaces directly (some other NoSQL databases do) so we
encode the namespace as a prefix of the keys under which we store a
application's data values. Many \name applications also use Redis'
built-in publish-subscribe mechanism to handle frequently changing
data.  For example, \boom{topology} publishes a notification to a
channel any of the keys in the topology namespace changes,
and \boom{routing} subscribes to this channel and updates its routing
configuration appropriately when it receives a notification that the
topology has changed.

\paragraph*{Applications}
Applications running on top of \name follow a common design pattern. Upon
startup, they connect with the \boom{nib} to retrieve any relevant
persistent state. 
The application then adds to, and retrieves from, the
persistent store any other necessary data depending on its
function. For example, \boom{topology} discovers and stores hosts,
switches, edges, and additional information about the topology in
the \boom{nib}, and when \boom{routing} starts up it reads this
information and then adds the least-cost paths to each destination.
During normal operation, applications are \emph{reactive}: they will
process events from the \boom{platform} and from other applications
(e.g., via the pub-sub mechanism). In response, they will make
changes to the NIB state and push out a new
NetKAT program via the \texttt{update} function on the \boom{platform}
nodes, which will update in the switches. 

\paragraph*{Update Coordinator}
Because \name has a distrib\-uted architecture, dynamic updates require
coordination between nodes. \name uses an update coordinator (or \boom{updc}) 
that manages interactions between
nodes during an update. We discuss these interactions in detail in the
next section.

\section{Dynamic updates with \Name}
\label{sec:updates}

\name's design supports dynamic updates by allowing important state to
persist in the NIB between versions while providing a way to transform
that state when required by an update. To ensure consistent semantics,
\name's \boom{updc} node organizes updates to the affected applications using a
simple protocol. This section describes this protocol, and then
describes some example updates that we have performed.


\subsection{Update protocol}\label{sec:upd_process}

To deploy an update, the operator provides \boom{updc} with the
following update specification:
\begin{itemize}
\item New versions of the affected applications' code
\item A state transformation function $\mu$ that maps the existing persistent 
state in affected namespaces into a format suited to the new
application versions.
\end{itemize}

As a convenience, the application programmer can write $\mu$ in a
domain-specific language (DSL) we developed for writing transformers
over JSON values (inspired by Kitsune's \emph{xfgen}
language~\cite{kitsune}), illustrated briefly in Sections~\ref{sec:fw}
and~\ref{sec:coord}. This language's programs are compiled to Python
code that takes an old JSON value and produces an updated version of
it.\footnote{While the programmer currently must write $\mu$,
automated assistance is also possible~\cite{kitsune, rubah}.}
Alternatively, the user can write $\mu$ using standard Python code.

Given the update specification, \boom{updc} then executes a
distributed protocol that steps through four distinct phases: (i)
application quiescence, (ii) code installation and state
transformation, (iii) application restart, and (iv) controller
resumption.

\paragraph*{1. Quiesce the applications}
\boom{updc} begins by signaling the applications designated for an update.
The applications complete any ongoing work and shut down,
signaling \boom{updc} they have done so. (A timeout is used to
forcibly shut down applications that do not exit gracefully.)  At the
same time, \boom{updc} sends the list of applications to
the \boom{platform}, which will temporarily suppress any rules updates
made by those applications, which could be stale.  Once all
applications have exited, and the \boom{platform} has indicated it has
begun blocking the rules, \name has reached \emph{quiescence}.

\paragraph*{2. Install the update in the \boom{nib}}
Next, \boom{updc} installs the administrator-provided $\mu$ functions
at the \boom{nib}.  The \boom{nib} verifies that these functions make
sense, e.g., that if the request is to update for
namespace \texttt{nodes} from versions \texttt{v3->v4}, then the
current \boom{nib} should contain namespace \texttt{nodes} at
version \texttt{v3}.  All transformations will be
applied \emph{lazily}, as part of in step 4.

\paragraph*{3. Restart the applications}
Now \boom{updc} begins the process of resuming operation.
\boom{updc} signals the new versions of the affected applications to start up.
These applications connect to the \boom{nib}, and the \boom{nib}
ensures that the applications' requested version matches the version
just installed in the \boom{nib}. The applications then retrieve
relevant state stored in the \boom{nib}, and compute and push the new
rules to the \boom{platform}.  The \boom{platform} receives and holds
the new rulesets. It will push them once it has received rules (or
otherwise been signaled) from \emph{all} of the updated applications,
to ensure that the rules were generated from consistent software
versions. Once the \boom{platform} has received rules from all updated
applications, it will remove the old rules previously created by the
updated applications and install the new rules on the switches.

\paragraph*{4. Resume operation}
At this point, the update is fully loaded and the applications proceed
as normal.  As the applications access data in the \boom{nib}, any
installed $\mu$ function is applied lazily.  In particular, when an
application queries a particular key, if that key's value has not yet
been transformed, the transformer is invoked at that time and the data
is updated. 

\bigskip{}

The rest of this section describes some example updates we have
implemented in \name for a stateful firewall, and for \boom{topology}
and \boom{routing} applications.

\subsection{Update example: Firewall}\label{sec:fw}

We developed three different versions of a stateful firewall, and
defined updates between them.

\begin{itemize}
\item \fwzero permits bidirectional flows between internal and
  external hosts as long as the connection is initiated from the
  inside. When the controller sees an outbound packet from internal
  host $S$ to external host $H$, it installs forwarding rules
  permitting communication between the two.

\item \fwone acts like \fwzero but only installs the rules
  permitting bidirectional flows after seeing returning traffic
  following an internal connection request. (It might do this to
  prevent attacks on the forwarding table originating from a
  compromised host within the network.)

\item \fwtwo adds to \fwone the ability to
time out connections (and uninstall their forwarding rules) after some
period of inactivity between the two hosts.
\end{itemize}



\fwzero defines a namespace \texttt{fw\_allowed} that keeps track of
connections initiated by trusted hosts, represented as JSON values:
\begin{verbatim}
{ "trusted_ip": "10.0.0.1", 
  "trusted_port": 3456, 
  "untrusted_ip": "10.0.0.2", 
  "untrusted_port": 80 }  
\end{verbatim}

Updating from \fwzero to \fwone requires the addition of a new
namespace, called \texttt{fw\_pending}; the keys in this namespace
track the internal hosts that have sent a packet to an external host
but have not heard back yet. 
Once the return packet is received, the host pair is moved to
the \texttt{fw\_allowed} namespace. For this update, no transformer
function is needed: all connections established under the \fwzero
regime can be allowed to persist, and new connections will go through
the two-step process.\footnote{We could also imagine moving all
currently approved connections to the pending list, but the resulting
removal of forwarding rules would be unnecessarily disruptive.}

Updating from \fwone to \fwtwo requires updating the data in the
\texttt{fw\_pending} and \texttt{fw\_allowed} namespaces, by adding
two fields to the JSON values they map to, \texttt{last\_count} and
\texttt{time\_created}, where the former counts the number of packets
exchanged between an internal and external host as of the time stored
in the latter. Every $N$ seconds (for some $N$, like 3), the firewall
application will query the \boom{nib} to see if the packet count has
changed. If so, it stores the new count and time. If not, it removes
the (actual or pending) route.

In our DSL we can express the transformation from \fwone to
\fwtwo data for the \texttt{fw\_allowed} namespace as follows:

\begin{verbatim}
  for fw_allowed:* ns_v0->ns_v1 {
    INIT ["last_count"] {$out = 0}
    INIT ["time_created"] {$out = time.time()} 
  };
\end{verbatim}

This states that for every key in the namespace, its corresponding
JSON value is updated from version \texttt{ns\_v0} (corresponding to
\fwone) to \texttt{ns\_v1} (corresponding to \fwtwo)
by adding two JSON fields. We can safely initialize the
\texttt{last\_count} field to $0$ because this is a lower bound on the
actual exchanged packets, and we can initialize \texttt{time\_created}
to the current time. Both values will be updated at the next timeout.
In general, our DSL can express transformations that involve adding,
renaming, deleting field names, modifying any data stored in the
fields, and also renaming the keys themselves. The DSL is detailed in
full in a separate work~\cite{kvolve} focusing on such updates.

The above code will be compiled to Python code that is stored (as a
string) in Redis and associated with the new version. The existing
data will be transformed as the new version accesses it via
the \boom{NIB} accessor API\@.
%
When the new version of the program retrieves connection information from the
\boom{nib}, the transformation would add the two new fields to the existing
JSON value shown earlier in this section:

\bigskip{}
\noindent
\begin{tabular}{ll}
\emph{key}: & \verb+fw_allowed:10.0.0.1_3456_10.0.0.2_80+ \\
\emph{value}: & \verb+{ "trusted_port": 3456, + \\
        & \verb+  "untrusted_port": 80, + \\
        & \verb+  "trusted_ip": "10.0.0.1",+\\
        & \verb+  "untrusted_ip": "10.0.0.2", +\\
        & \verb+  "last_count": 0, +\\
        & \verb+  "time_created": 1426167581.566535 }+\\
\end{tabular}





\subsection{Coordination: Routing and Topology}
\label{sec:coord}

In the above example, the firewall is storing its own data in the \boom{nib}
with no intention of sharing it with any other applications. As such, we
could have killed the application, installed the update, and
started the new version. However, when multiple applications share the
same data and its format changes in a backward-in\-com\-pat\-ible manner,
then it's critical that we employ the update protocol described in
Section~\ref{sec:upd_process}, which gracefully coordinates the
updates to applications with shared data.



As an example coordinated update, recall from
Section~\ref{sec:boombox} that our \boom{routing} and \boom{topology}
applications share topology information stored in the \boom{nib}. In
its first version, \boom{topology} merely stores information about hosts,
switches, and the links that connect them. The \boom{routing} application
computes per-source/destination routes, assuming nothing about the
capacity or usage of links. In the next version,
\boom{topology} regularly queries the switches for port statistics and
stores the moving average of each link's bitrate in the \boom{nib}.
This information is then used by
\boom{routing} when computing paths. The result should be better load
balancing when multiple paths exist, between hosts.

Updating from the first to the second version in \name requires
adding a field to the JSON object for edges, to add the measured
bitrate. 
The transformer $\mu$ simply initializes this field to 1,
indicating the default value for traffic on the link as follows:
\begin{verbatim}
  for edge:* ns_v0->ns_v1 {
    INIT ["weight"] {$out = 1} 
  };
\end{verbatim}
As such, the initial run of
the routing algorithm will reproduce the existing routes because all initial
values will be the same, ensuring
stability. Subsequent \boom{routing} computations will account for and store the
added usage information and thus better balance the routes.


\section{Experiments and Evaluation}
\label{sec:exps}

In this section, we report the results of experiments where we dynamically
update several canonical SDN applications: a load balancer, a firewall, and a
routing application. We implement three dynamic update mechanisms: state
transfer using \name, simple restart, and record and replay. In all cases,
state transfer is fast and disruption-free, whereas the other techniques cause
a variety of problems, from network churn to dropped connections.
We ran all experiments using Mininet HiFi~\cite{hifi}, on an Intel(R) Core(TM)
i5-4250U CPU @ 1.30GHz with 8GB RAM. We report the average of 10 trials.

\subsection{Firewall}\label{sec:firewall}

\begin{figure}
 \centering
   \begin{subfigure}{.39\textwidth}
   \begin{tikzpicture}
   \node{\pgfimage[width=\textwidth]{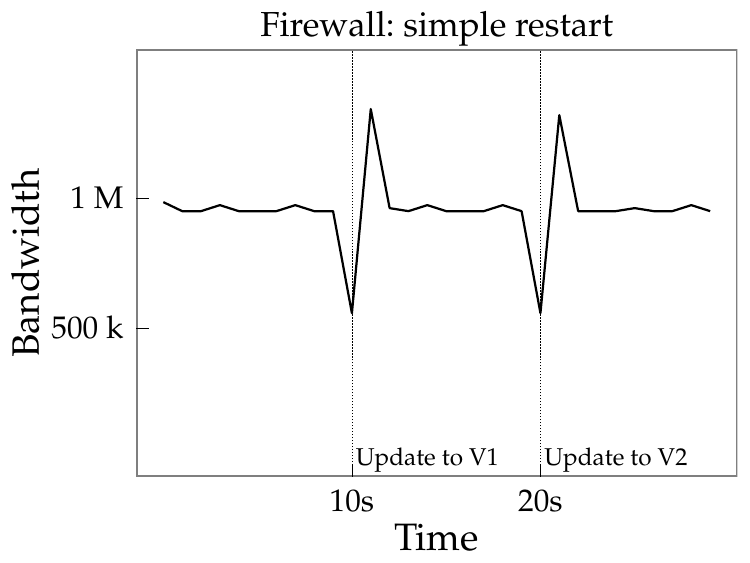}};
   \end{tikzpicture}
   \end{subfigure}
   \begin{subfigure}{.39\textwidth}
   \begin{tikzpicture}
   \node{\pgfimage[width=\textwidth]{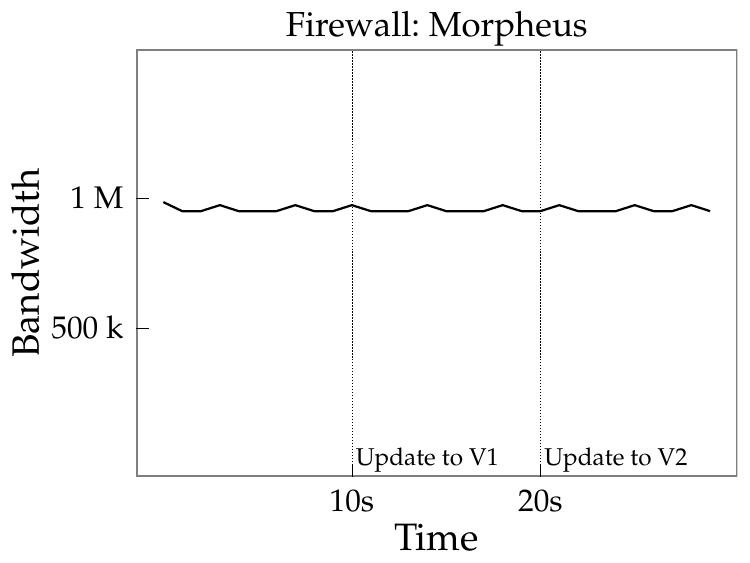}};
   \end{tikzpicture}
   \end{subfigure}
 \caption{Firewall Update}
 \label{fig:firewall_update}
\end{figure}

Figure~\ref{fig:firewall_update} illustrates a dynamic update to the
firewall, described in Section~\ref{sec:fw}, from \fwzero to \fwone and then to
\fwtwo. The figure shows the result of simple restart (where all data is stored
in memory and lost on restart) and state transfer (where data is stored in the
\boom{NIB}). We do not depict record and replay, which happens to perform
as well as state transfer for this example (as per Section~\ref{sec:rr}).

For the experiment, we used a single switch with two ports (with 1 MBPS
bandwidth) connected to two hosts. One host is designated the client inside the
firewall and the other is the server outside the firewall. Using iperf, we
establish a TCP connection from the client to the server. The figure plots the
bandwidth reported by iperf over time. In both experiments, we update to \fwone
after 10 seconds and \fwtwo after 20 seconds.

Using simple restart, the figure shows that bandwidth drops significantly during updates.
This is unsurprising, since a newly started firewall doesn't remember existing
connections. Therefore, \fwone and \fwtwo first block all packets from the server to the client,
until the client sends a packet, which restores firewall state. In contrast,
\name doesn't drop any packets because state is seamlessly transformed from
one version to the next.

\subsection{Routing and Topology}\label{sec:routtop}

 \begin{figure}
  \centering
    \begin{subfigure}{.39\textwidth}
    \begin{tikzpicture}
    \node{\pgfimage[width=\textwidth]{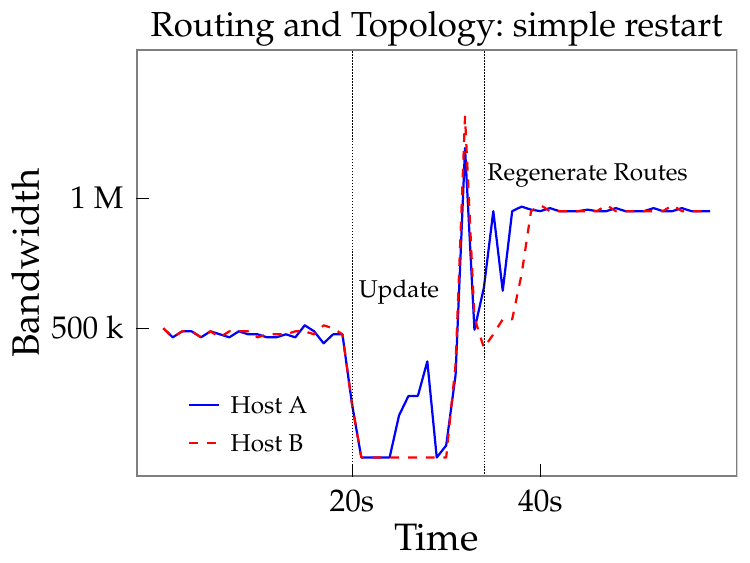}};
    \end{tikzpicture}
    \end{subfigure}
    \begin{subfigure}{.39\textwidth}
    \begin{tikzpicture}
    \node{\pgfimage[width=\textwidth]{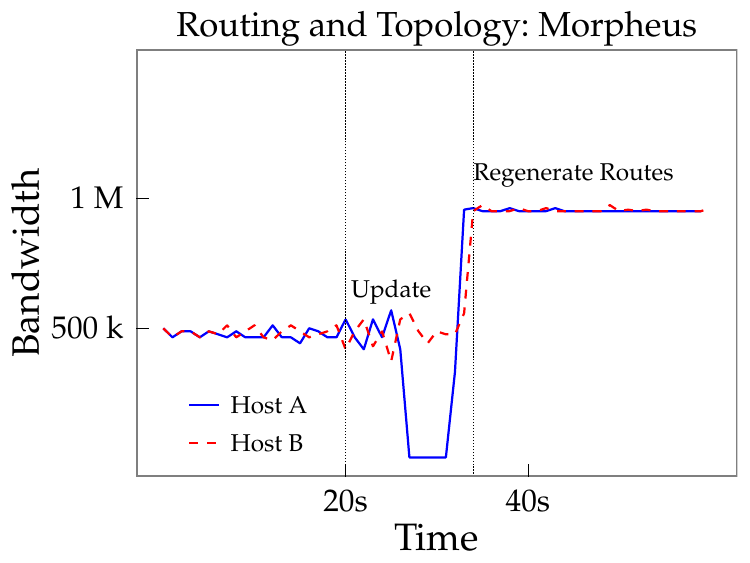}};
    \end{tikzpicture}
    \end{subfigure}
  \caption{Routing and Topology Discovery Update}
  \label{fig:topo_upd}
 \end{figure}

Figure~\ref{fig:topo_upd} shows the effect of updating routing and topology
applications (described in section~\ref{sec:coord}), where the initial version
uses shortest paths and the final version takes current usage into
account. The experiment uses four
switches connected in a diamond-shaped topology 
with a client and server on either end. Therefore, there are two paths of equal
length through the network. The client establishes two iperf TCP connections to
the server.

Initially, both connections are routed along the same path because the
first version of  \boom{topology} and \boom{routing} pick the same
shortest path. The links along the path are 1MBPS, therefore each connection
gets 500K\-BPS by fair-sharing.
After 20 seconds elapse, we update both applications: the new version of
\boom{topology} stores link-utilization information in the NIB and the new
version of \boom{routing} using this information to balance traffic across
links. After the update, each connection should be mapped to a unique path,
thus increasing link utilization and the bandwidth reported by iperf.

Using simple restart, both connections are disrupted for 10 seconds, which is
how long \boom{topology} takes to learn the network topology. Until
the topology is learned, routing can't route traffic for either connection.
\name is much less disruptive. Since the state transfer function preserves
topology information, the new \boom{routing} module maps each connection to a
unique path. The connection that is not moved (Host B) suffers no disruption
and gracefully jumps to use 1MBPS bandwidth. The connection that is moved (Host
A) is briefly disrupted as several switch tables are updated. Even this
disruption could be avoided using a consistent update~\cite{consistentupdate}.

\begin{table}[t]
\centering
\footnotesize
\def\arraystretch{1.1}
\begin{tabular}{@{}cccccc@{}}\toprule
\multirow{2}{*}{\emph{start}} & \emph{apps}  & \emph{restart} &  \emph{rout} & \emph{topo}  & \emph{platform} \\
& \emph{exit}  & \emph{begins}  &  \emph{push} & \emph{push}  & \emph{resume}  \\
\midrule
0.00s   &   0.05s   &   0.11s   &   1.67s   &   1.68s   &   1.70s  \\
\bottomrule
\end{tabular}
\caption{Update Quiescence Times for \boom{Topology} and \boom{Routing} (Median of 11 trials)}
\label{table:qui}
\end{table}

Table~\ref{table:qui} breaks down the time to run the update protocol
for this update.
 It takes .05s for both \boom{topology}
and \boom{routing} to receive the signal to exit at their quiescent points and
shut down, and for the \boom{platform} to also receive the signal and pause. At
.11s, both applications restart, begin pulling from the \boom{nib}, and begin
performing computations. At 1.67s and 1.68s respectively, the \boom{routing}
and \boom{topology} applications send their newly computed rules to the \boom{platform}. The
\boom{platform} holds on to the rules until it ensures it has received the rules
from both apps, and then \boom{platform} pushes both sets of rules to the switches
and unpauses. This entire process takes 1.70s, with most of the time
taken by simply restarting the application (as would be required in
the simple case anyway). In general, the amount of time
to update multiple applications safely will vary based on number of
applications, the amount of state to restore, and the type computations to be
performed to generate the rules, but the overhead (compared to a
restart) seems acceptable.

\subsection{Load Balancer}\label{sec:loadbal}
\begin{figure*}
\begin{subfigure}{.33\textwidth}
\begin{tikzpicture}
\node{\pgfimage[width=\textwidth]{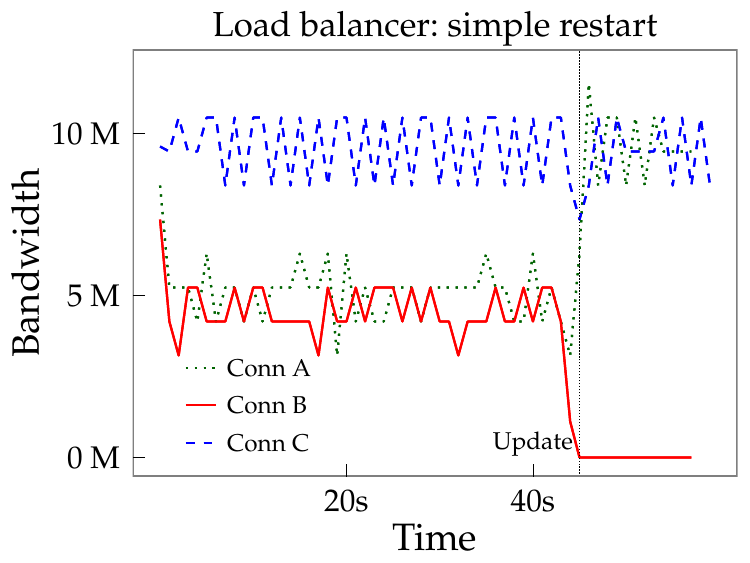}};
\end{tikzpicture}
\end{subfigure}
\begin{subfigure}{.33\textwidth}
\begin{tikzpicture}
\node{\pgfimage[width=\textwidth]{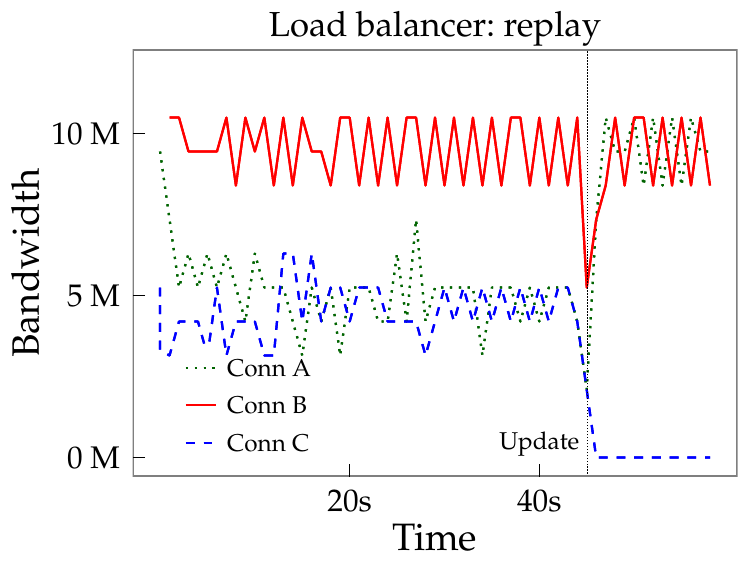}};
\end{tikzpicture}
\end{subfigure}
\begin{subfigure}{.33\textwidth}
\begin{tikzpicture}
\node{\pgfimage[width=\textwidth]{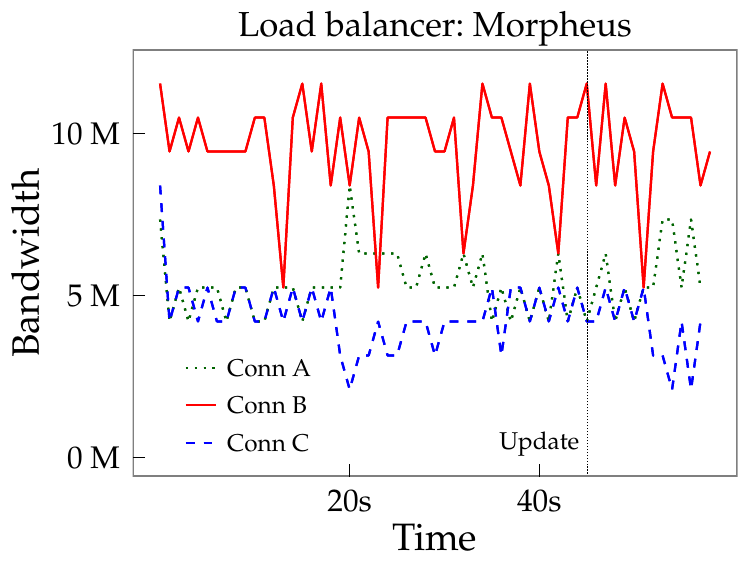}};
\end{tikzpicture}
\end{subfigure}
\caption{Load Balancer Results}
 \label{fig:lb}
\end{figure*}

Figure~\ref{fig:lb} shows the effect of updating a load-balancer that maps
incoming connections to a set of server replicas.  For this experiment, in
addition to the simple restart and \name experiments, we also report the
behavior of
record-and-replay which consists of recording the packet-in events and
replaying them after restart.  After 40 seconds, we bring an additional server
online and update the application to also map connections to this server.
To avoid disconnecting clients, existing connections should not be
moved after the update. 

As shown in the figure, both simple restart and record-and-replay
cause disconnections, whereas state transfer causes no
disruptions, since the state is preserved. As discussed in Section~\ref{sec:rr}, 
replaying the recorded packet-ins will cause the three connections to be evenly
distributed across the three servers.  Similarly, for the simple restart, the
connections will be evenly distributed when the clients attempt to reconnect.
Therefore, one connection is erroneously mapped to the new server mid-stream,
which terminates the connection.

\subsection{Programmer Effort}\label{sec:effort}




Starting from a \name application/service, there are two main additional
tasks required to enable dynamic update: writing code to quiesce an
application prior to an update, and writing a $\mu$
transformer function to change the state. In this subsection we
discuss both tasks, showing that both are straightforward.

\paragraph*{Quiescence} The application
developer must write code to check for notifications from the
\boom{nib} that an update is available, and if so to 
complete any outstanding tasks and gracefully exit. These
tasks would include storing any additional state in the \boom{nib}
and/or notifying external parties.
For all of our examples, this work was quite simple, amounting to 
8 lines of code. 

\paragraph*{Transforming the state} Writing the function $\mu$ to
transform the state was also straightforward.
For \boom{firewall}, as described in Section~\ref{sec:fw}, we wrote 4 lines of
DSL code to initialize new fields to desired values so that the fields could be
read with the correct data.  Similarly for our applications \boom{topology} and
\boom{routing}, as described in Section~\ref{sec:coord}, we wrote 3 lines of
DSL code to initialize the weight field to a default value.  For the \boom{load
balancer}, no $\mu$ function was necessary, as no state was transformed, only
directly transferred to the new version of the program.

We also looked at
the application histories of other controllers to get a sense of how
involved writing a $\mu$ function might be for updates that occur
``in the wild.''  In particular, we looked at 
GitHub commits from 2012--2014 for OpenDaylight~\cite{odl} and POX~\cite{pox}
applications.  We examined applications such as a host tracker, a topology
manager, a Dijkstra router, an L2 learning switch, 
a NAT, and a MAC blocker.
Several of the application changes consisted only of updates to the application
logic, such as multiple changes to POX's IP load balancer in 2013. For them, no
$\mu$ would be necessary.  We also found that many of the application changes
involved adding state, or making small changes to existing state.
For example, an update to OpenDaylight's host tracker on November 18, 2013
converted the representation of an \texttt{InetAddress} to a \texttt{IHostId}
to allow for more flexibility and to store some additional state such as the
data layer address.  To create this update, the administrator would write $\mu$
to initialize the data layer address for all stored hosts, if known, or add
some dummy value to indicate that the data layer address was not known.  An
update to POX's host tracker on June 2, 2013 added two booleans to the state to
indicate if the host tracker should install flows or should suppress
ARP replies.  To create this update, the administrator would write 
$\mu$ to initialize these to \texttt{True} in the \boom{nib}.  To sum up, 
while the size of $\mu$ scales with the size of the change in state
being made, in practice, we found that the effort to write $\mu$ is minimal.

\section{Related work}\label{sec:related_work}

\name represents the first general-purpose solution to the problem of
dynamically updating SDN controllers (and by extension, updating the
networks they manage). We argued this point extensively in
Section~\ref{sec:problem}, specifically comparing to alternative
techniques involving controller restarts and record and replay
(exemplified by the HotSwap system~\cite{hotswap}). In this section we
provide comparison to other work that provides some solution
to the dynamic update problem.

\paragraph*{Graceful control-plane updates} Several previous works have looked at the problem of
updating control-plane software. In-Service Software Upgrades (ISSU)~\cite{CiscoISSU,JuniperISSU}
minimize control-plane downtime in high-end routers upon an OS upgrade by installing the new control
software in parallel with the old one, on different blade and synchronizing the state automatically.
Other research proposals go even further and allow other routers to respond correctly to topology
changes that affect packet forwarding, while waiting for a peer to restart its control
plane~\cite{shaikh_graceful_shutdown_of_ospf_02, ShaikhGracefulShutdownMultipleRoutersTON06}.
In general, most routing protocols have mechanisms to rebuild their state when the control
software (re)starts (cf. \cite{rfc3623,rfc4724}), e.g., by querying
the state of neighboring routers. 

The key difference between these works and \name is that \name aims to
support unanticipated, semantic changes to control-plane software,
possibly necessitating a change in state representation, whereas ISSU
and normal routing protocols cannot.\footnote{Cisco only supports ISSU
  between releases within a rolling 18-month
  window~\cite{cisco_issu_problems}. Outside of this window, a
  hard-reset of the control-plane has to be done.} In addition, \name
is general-purpose (due to its focus on SDN), and not tied to a
specific protocol design (e.g., a routing protocol).

\paragraph*{Distributed Controllers} Distributed SDN controller
architectures such as Onix~\cite{Koponen:2010th},
Hyperflow~\cite{hyperflow}, ONOS~\cite{onos} or Ravana~\cite{ravana}
can create new controller instances and synchronize state among them
using a consistent store. \name's distributed design is inspired by
the design of these controllers, which aim to provide scalability,
fault-tolerance and reliability, and can support simple updates in
which the shared state is unchanged between versions (and/or is
backward compatible).
However, to the best of our knowledge
these systems have not looked closely at the controller upgrade problem when
(parts of) the control program itself must be upgraded in a
semantics-changing manner, especially when
the new controller may use different data structures and algorithms
than the old one. \name handles this situation using the update
protocol defined in Section~\ref{sec:updates}, which quiesces
the controller instances, initiates a transformation of the shared store's
data according to the programmer's specification (if needed), and then
starts the new controller versions. We believe this same approach
could be applied to these distributed controllers as well.


\paragraph*{Dynamic Software Upgrades} The approach we take in
\name is inspired by a line of work on \emph{dynamic software
  updating} (DSU)~\cite{HicksNettles03,neamtiu09stump,kitsune,rubah},
which advocates the same basic approach: pause a program threads at
quiescent points, transform and transfer state into the new version of
the program, and resume execution in the updated version of a program.
Most prior DSU work has focused on updating a running process
(``bringing the new code to the old (but transformed) data'') whereas
for \name the same effect is achieved by starting a new process with
the relevant state (``bringing the old (but transformed) data to the
new code''). We used our KVolve~\cite{kvolve} system for dynamically
evolve Redis databases in our implementation of \name.
While prior DSU work has considered the problem
updating network software generally~\cite{SegalF93} (including for
``active'' networks~\cite{HicksN00}), ours is the first to apply a
general-purpose solution to (distributed) software-defined network
controllers in particular.

\section{Conclusions}
\label{sec:conclusion}

This paper has proposed \emph{dynamic update by state transfer}
as a general-purpose approach to dynamically update
software-defined network controllers.
The approach works by providing direct access to the
relevant state in the running controller, and initializing the new
controller's state as function of the existing state. This approach is
in contrast to alternatives that attempt to automatically reproduce
the relevant state, but may not always succeed. We implemented
the approach as part of \name, a new SDN controller whose design is
inspired by industrial-style controllers. \name provides means to specify
transformations in a persistent store, and employs an
update coordination protocol to safely deploy the transformation.
Experiments with \name show that dynamic update by state transfer is
both natural and effective: it supports seamless updates to live
networks at low overhead and little programmer effort, while prior
approaches would result in disruption, incorrect behavior, or both.


{\small
\balance
\bibliographystyle{abbrv}
\bibliography{sigproc} 
}
\end{document}